\documentclass[twocolumn,twoside,slac_two]{revtex4}
\usepackage{graphicx}
\usepackage{fancyhdr}
\usepackage{hyperref}
\hypersetup{
    colorlinks=true,
    urlcolor=magenta,
}
\usepackage{xspace}
\newcommand{\Xm}{\ensuremath{X_{\rm max}}\xspace}%{$X_{\rm{max}}\,\,$}

\begin{document}

\title{Review on extragalactic cosmic rays detection}

\author{Mariangela Settimo}
\affiliation{Laboratoire de Physique Nucl\'eaire et de Hautes Energies, LPNHE, IN2P3/CNRS Paris }

\begin{abstract} 
The understanding of the nature of ultra-high-energy cosmic rays is one of the most intriguing open questions for current and future observatories. These particles are expected to be accelerated in extragalactic sources. Because of their low flux, their detection is done indirectly observing extensive air-showers, with ground-based experiments. A review of the detection technique and the most recent results is presented with a look to the perspectives for the future. 
\end{abstract}

\maketitle

\thispagestyle{fancy}

\section{Introduction}

The origin and the nature of ultra-high-energy cosmic rays (UHECR), above about 10$^{18}$~eV are among the long-standing open questions in astroparticle physics. At so high energy, cosmic rays are not anymore confined in the galaxy and they are expected to be of extragalactic origin. 
Their astrophysical sources have not yet been identified and the distribution of their arrival directions is showing a very weak anisotropy level, challenging the possibility of doing astronomy with the highest energetic cosmic rays. A detailed review of the cosmic-ray anisotropy studies has been given at this conference~\cite{Deligny}.  The energy spectrum is a valuable tool to get hints on the changes in the production sites or in the nature of cosmic rays.  Two spectral features have been observed at ultra-high energy (UHE): a hardening around 10$^{18.5}$~eV, the ``ankle'', and a flux suppression above 10$^{19.5}$~eV~\cite{HiResSpectrum,AugerSpectrum2015,TASpectrum2015} . 
Despite the challenging astrophysical interpretation, the ankle has been historically linked to the transition from a steep galactic component to a flat extragalactic one. In the so-called ``dip model",  the ankle is interpreted as due to the pair-production by protons interacting with the cosmic microwave background (CMB) and the transition from the galactic to the extragalactic component is expected to occur at energies around 10$^{17}$~eV (see e.g.~\cite{DipModel,TaylorReview}). 
Similarly the interpretation of the flux suppression at the highest energies is still controversial~\cite{GZK1,GZK2,Allard,Allard2}. Two possible scenarios can describe the current data: (i) In the ``propagation scenario" the flux suppression is due to the photo-pion production by protons above 10$^{19.5}$~eV interacting with the CMB (the so-called ``GZK effect"), or more generally to the photo-disintegration of nuclei; (ii) In the ``source exhaustion" scenario, cosmic rays can be accelerated up to a maximum energy proportional to their charge, $Z$: $$E_{\rm{max}} \propto Z\times E_{\rm{max}}^{\rm{p}}$$ 
with $E_{\rm{max}}^{\rm{p}}$ the maximum acceleration energy for proton primaries.  
In the latter scenario, the flux suppression is thus a direct consequence of the flux cut-off at the source and the chemical composition is expected becoming heavier with increasing energy. 
\begin{figure}[b]
\centering
\includegraphics[width=0.45\textwidth]{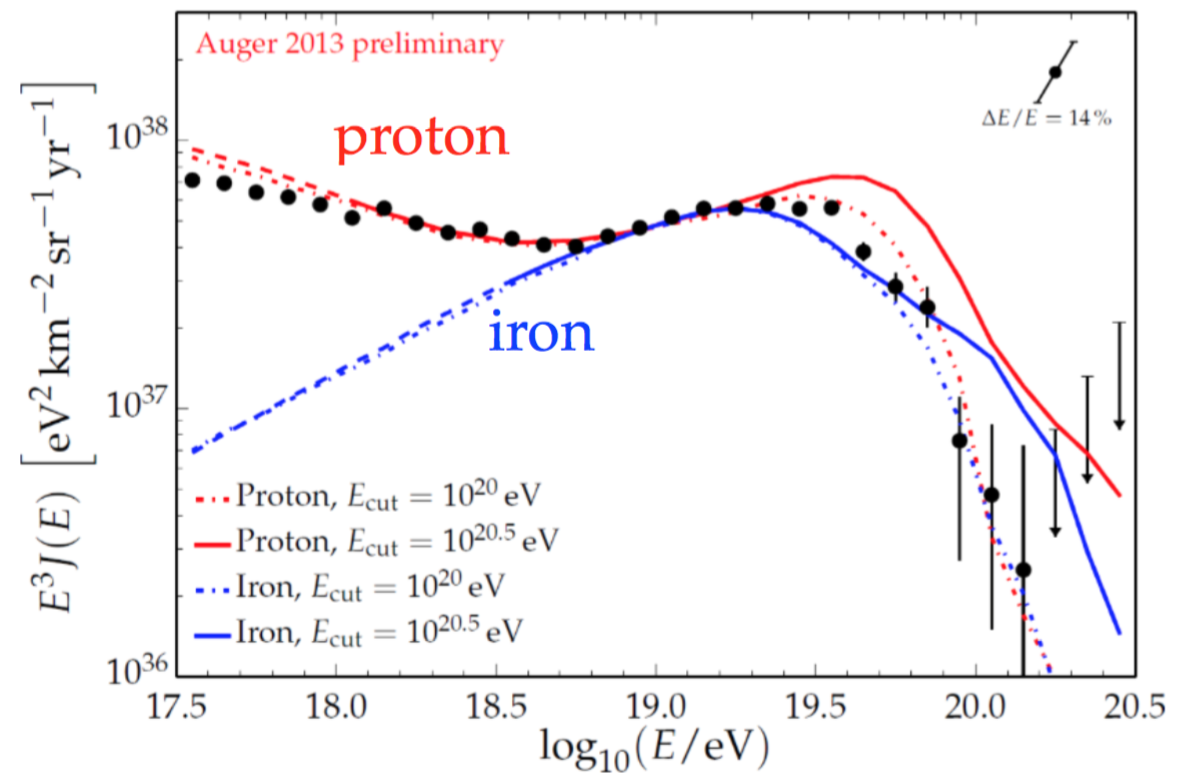}
\includegraphics[width=0.45\textwidth]{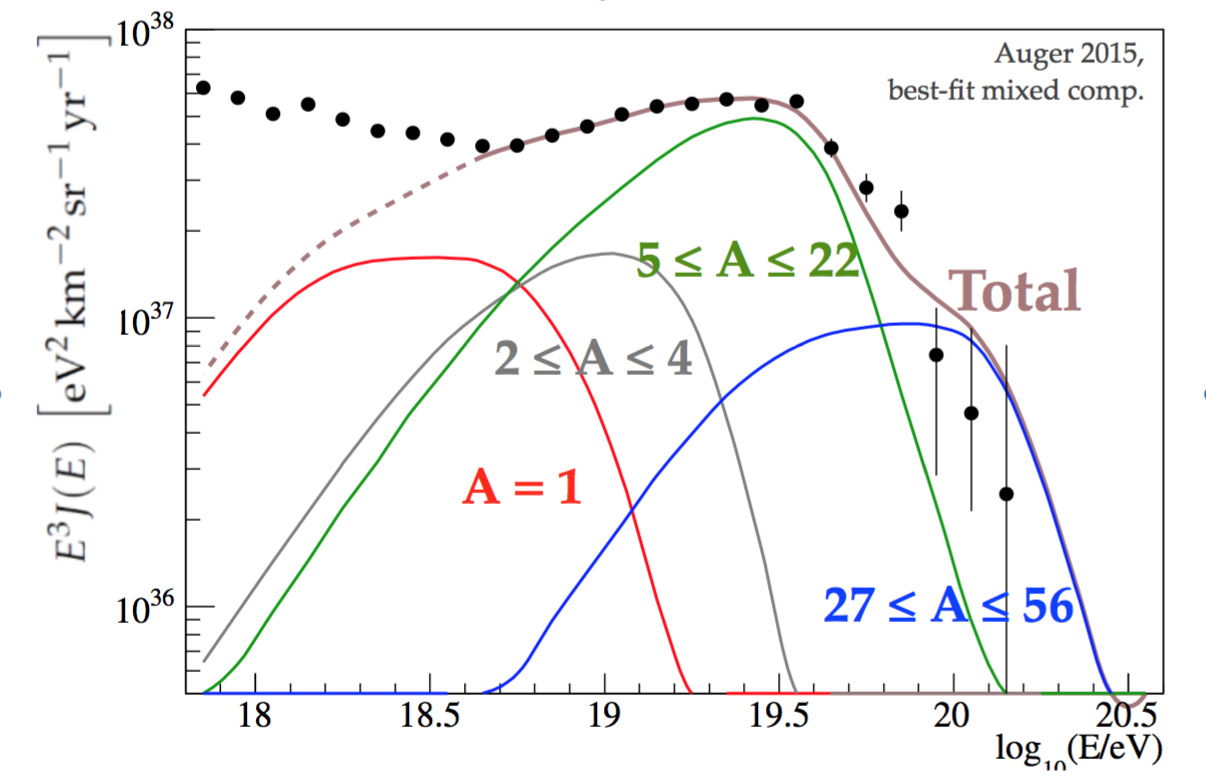}
\caption{Possible interpretations of the flux suppression observed at energies above 10$^{19.5}$~eV. Top:  the ``propagation scenario"  (GZK effect) for proton and iron primary cosmic-rays and two different maximum energy thresholds; Bottom: best fit of the spectrum assuming the ``source-exhaustion" scenario with a 4-components mixed composition~\cite{AugerPrime}. 
 \label{fig:spectra}}
\end{figure}

Fig.~\ref{fig:spectra} shows two examples of energy spectrum fit for the "propagation" (top, proton and iron cases) and the ``source exhaustion" (bottom) scenarios~\cite{AugerPrime}. A discrimination between these two scenarios based only on the energy spectrum is not possible and the mass composition provides crucial complementary informations. 

In addition, the observation of a diffuse flux of cosmogenic photons and neutrinos, from the decay of pions produced in the GZK effect, would constitute an independent proof in favour of the propagation scenario. 

The experimental results on the energy spectrum, the mass composition and the cosmogenic secondaries are summarised in the next sections. The constraints on the hadronic interaction models are also discussed as they constitute an important ingredient in the interpretation of the experimental results. The anisotropy searches at large and small scales, a key element of the UHECRs puzzle, are discussed in detail in another contribution at this conference~\cite{Deligny}. 

\section{Detection techniques}

Given their low flux, of about a few particles per km$^2$ per year,  UHECRs cannot be detected with satellite- or space-born experiments. 
On the other hand, the secondary-particles cascades that UHECRs induce in atmosphere are on average enough deep and with a  sufficiently large footprint to be detected at ground level. 
A few quantities, such as the evolution of the electromagnetic cascade, the density of the air-shower particles at the ground and the muonic content, are needed to derive the properties of the primary particle hitting the top of the atmosphere (i.e., energy, mass composition, arrival direction).  

Two detection techniques are extensively employed: the observation of the fluorescence light isotropically emitted by the nitrogen molecules during the passage of the air-shower in the atmosphere by means of telescopes (FD) and the measurement of the particle density at the ground by an array of surface detectors (SD). 
The FD directly accesses the longitudinal profile of the air-shower (the energy deposit, $dE/dX$, as a function of the atmospheric depth $X$) whose integral provides a measurement of the calorimetric energy of the primary particle.  
The total energy is then obtained correcting for the invisible energy carried by penetrating particles (mostly neutrinos and muons), that amounts to around 10-15\% for nuclear primaries and only weakly depends on the primary mass and on the hadronic interaction models. 
Moreover,  FD allows a direct measurement of the depth (\Xm) at which the air-shower reaches its maximum development that is a well known mass composition sensitive parameter.  
However, this detector only operates during clear and moonless nights, limiting the duty cycle to about 15\%. In addition the atmospheric scattering and absorption, which influence the propagation of the light from the axis to the telescope, fix the parameters for the detector design and require continuously running atmospheric monitoring systems for a precise knowledge of the atmospheric condition and the aerosol content. 

\begin{figure}
\centering
\includegraphics[width=0.45\textwidth]{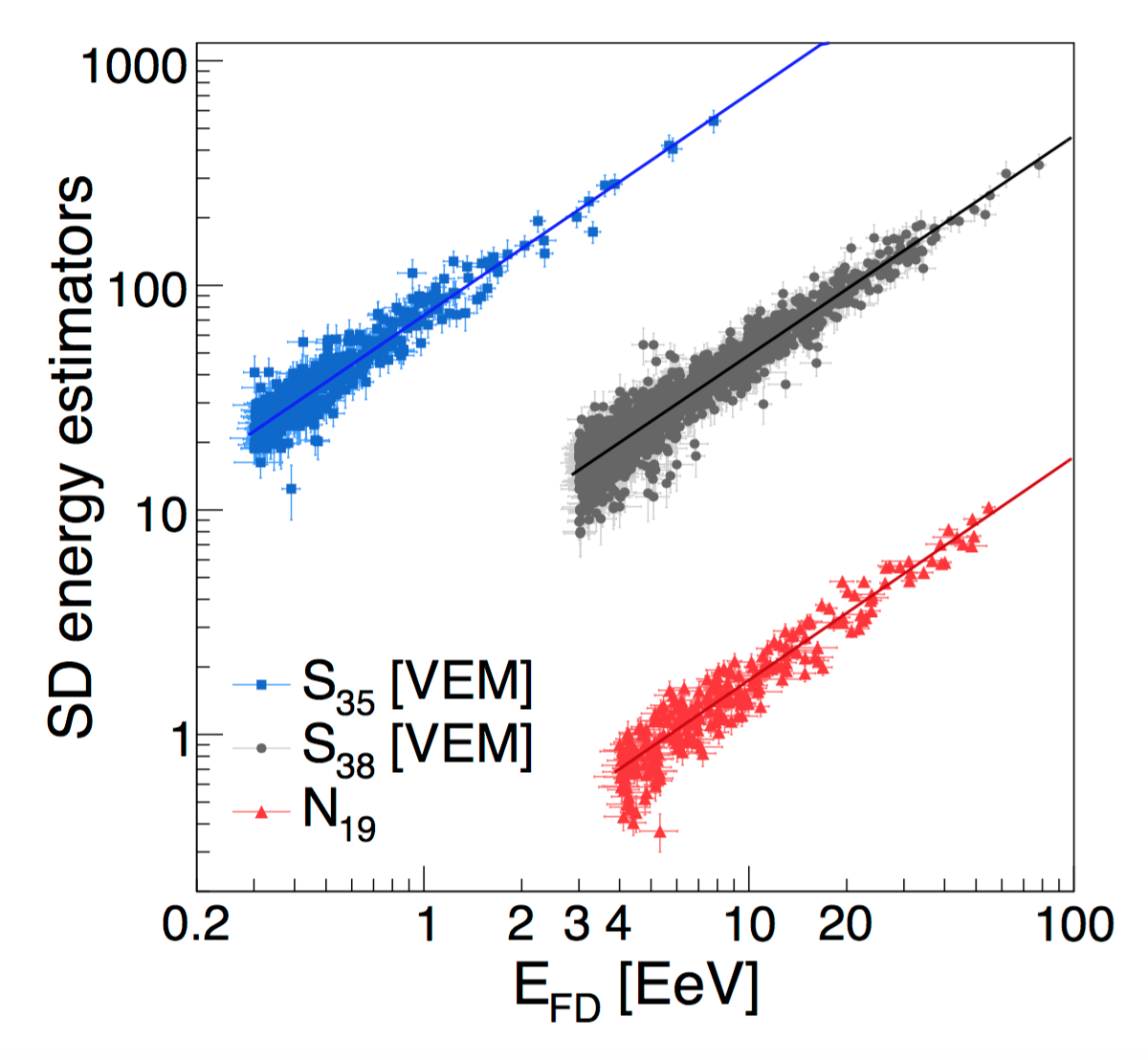}
\caption{Data-driven energy calibration using the hybrid detection mode (here for the Pierre Auger Observatory).  The SD energy estimator, for three different datasets, is calibrated with the calorimetric energy measured by FD for the sub-sample of ``hybrid events"~\cite{AugerSpectrum2015}. \label{fig:calib}}
\end{figure}

The use of an array of particle detectors at ground, as water-Cherenkov stations or scintillators, allows for a 100\% duty cycle and the coverage of huge surfaces. 
The surface detectors (SD) sample the particle density at the ground as a function of the radial distance (lateral distribution function, LDF). The interpolated signal ($S_{\rm{ropt}}$) at a reference distance $\rm{ropt}$ from the shower axis~\cite{Watson} is adopted as energy estimator. When the SD operates alone, the calibration of the energy estimator is based on simulations 
To get rid of the model dependence, a ``hybrid detection" mode has been conceived and pioneered by the Pierre Auger Observatory almost 20 years ago. It is based on the combined use of  two independent techniques: events observed at the same time by the SD and the FD, named ``hybrids", are used to perform an FD-data driven calibration of the SD energy estimator. An example of the calibration procedure is shown in Fig.~\ref{fig:calib} for  the three different datasets used by the Pierre Auger Observatory, the biggest experiment currently in operation for UHECRs.  

\begin{figure}[!t]
\centering
\includegraphics[width=0.45\textwidth]{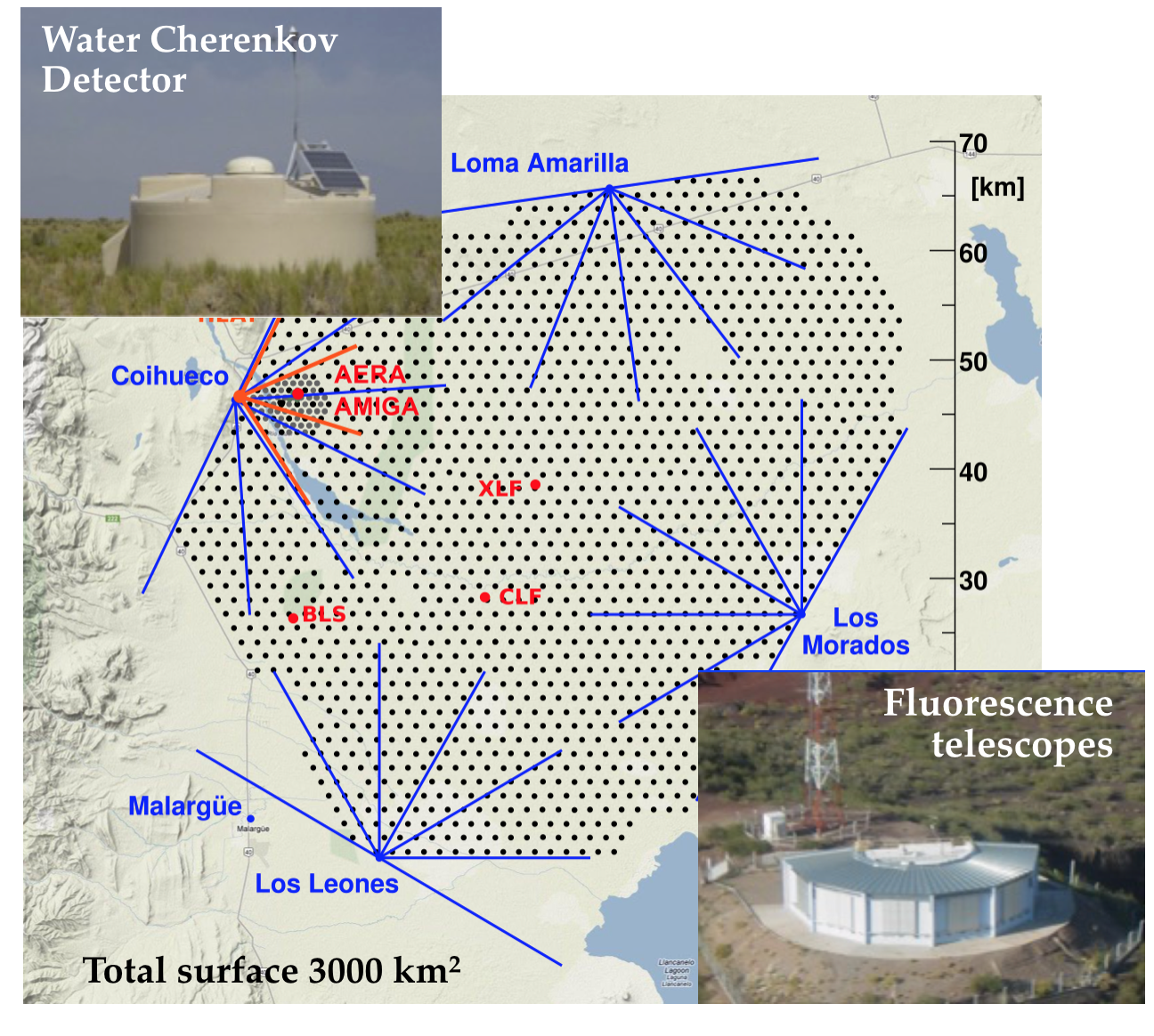}
\includegraphics[width=0.45\textwidth]{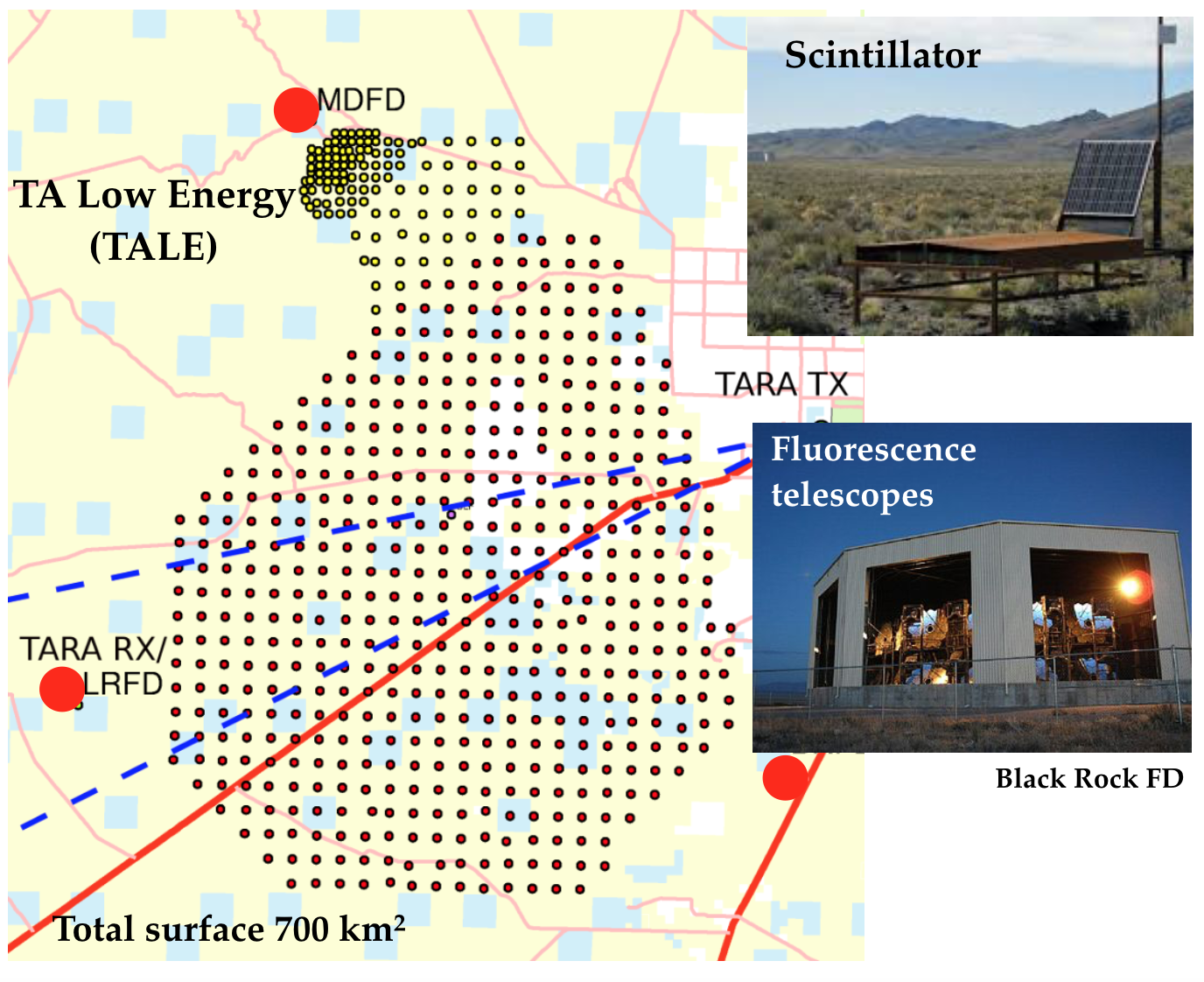}
\caption{Map of the Pierre Auger Observatory~\cite{AugerNIM2015} (top) and the Telescope Array~\cite{TelescopeArray} (bottom) with their main detectors, the fluorescence telescopes and the water-Cherenkov or scintillator stations. 
 \label{fig:observatories}}
\end{figure}

In the past years, several efforts have been conducted in order to develop new detection techniques that overcome the statistical limitation of the fluorescence telescopes while preserving the possibility of measuring the longitudinal profile (e.g., \cite{euso,radio,GIGAS,TARA}). Among them, the radio technique has been proved to work well in the energy region around 10$^{17} - 10^{18}$ using arrays of antenna with a spacing of few hundreds meters to sample the radio signal lateral distribution~\cite{LOFAR}.\\

Two experiments are currently taking data, one in each hemisphere, the Telescope Array in Utah (USA) and the Pierre Auger Observatory in Argentina. The Telescope Array consists of an array of 3~m$^2$ scintillators deployed over a surface of about 700~km$^2$ and 38 fluorescence telescopes at the edges of the array~\cite{TelescopeArray}. 
The Pierre Auger Observatory comprises an array of 1660 water-Cherenkov stations, each one with a surface of 10~m$^2$ and a height of 1.2~m, covering an area of 3000~km$^2$ and overlooked by 24 fluorescence telescopes grouped in 4 buildings~\cite{AugerNIM2015}. Additional surface stations, deployed in a dense array, and high-elevation telescopes are installed in both observatories to enhance the detection of low-energy cosmic-rays, below 10$^{18}$~eV. 
Fig.~\ref{fig:observatories} shows a sketch of the array configuration and of the two main detectors of the Pierre Auger Observatory (top) and the Telescope Array (bottom). 
These experiments, which are taking data since more than 10 years, have achieved many interesting and unexpected results, some of them reported in the next sections. More comprehensive reviews can be found in~\cite{TAICRC2015,AugerICRC2015}. 

\section{Energy spectrum}
One of the most relevant results is the measurement of the energy spectrum and the precise determination  of  its spectral features (see Fig.~\ref{fig:spectra}). 
The spectra measured by the Auger and by the Telescope Array observatories are in good agreement within the systematic uncertainties, quoted as 14\% for the Pierre Auger Observatory and 21\% for Telescope Array. The main sources of systematic uncertainties are related to the atmospheric knowledge, the detector calibrations and the fluorescence yield.  
Even if compatible within uncertainties, a discrepancy in the cut-off region remains and it is not clearly understood. 
The possibility that it is originated by a different sky in the northern and southern hemispheres has been tested evaluating the declination ($\delta$) dependence of the energy spectrum. The Auger Collaboration has not found any significant change in the flux measured in four declination bands and the only evident variations are compatible with the a dipolar modulation of the flux found at energies above $8\times10^{18}$~eV~\cite{Deligny}.
The Telescope Array has shown some preliminary hints of a change of the spectral cut-off when selecting events with $\delta < 26^\circ$ and $\delta > 26^\circ$ but a firm conclusion cannot be drawn with the current statistics of events~\cite{TASpectrum2015}.

\begin{figure}
\centering
\includegraphics[width=0.51\textwidth]{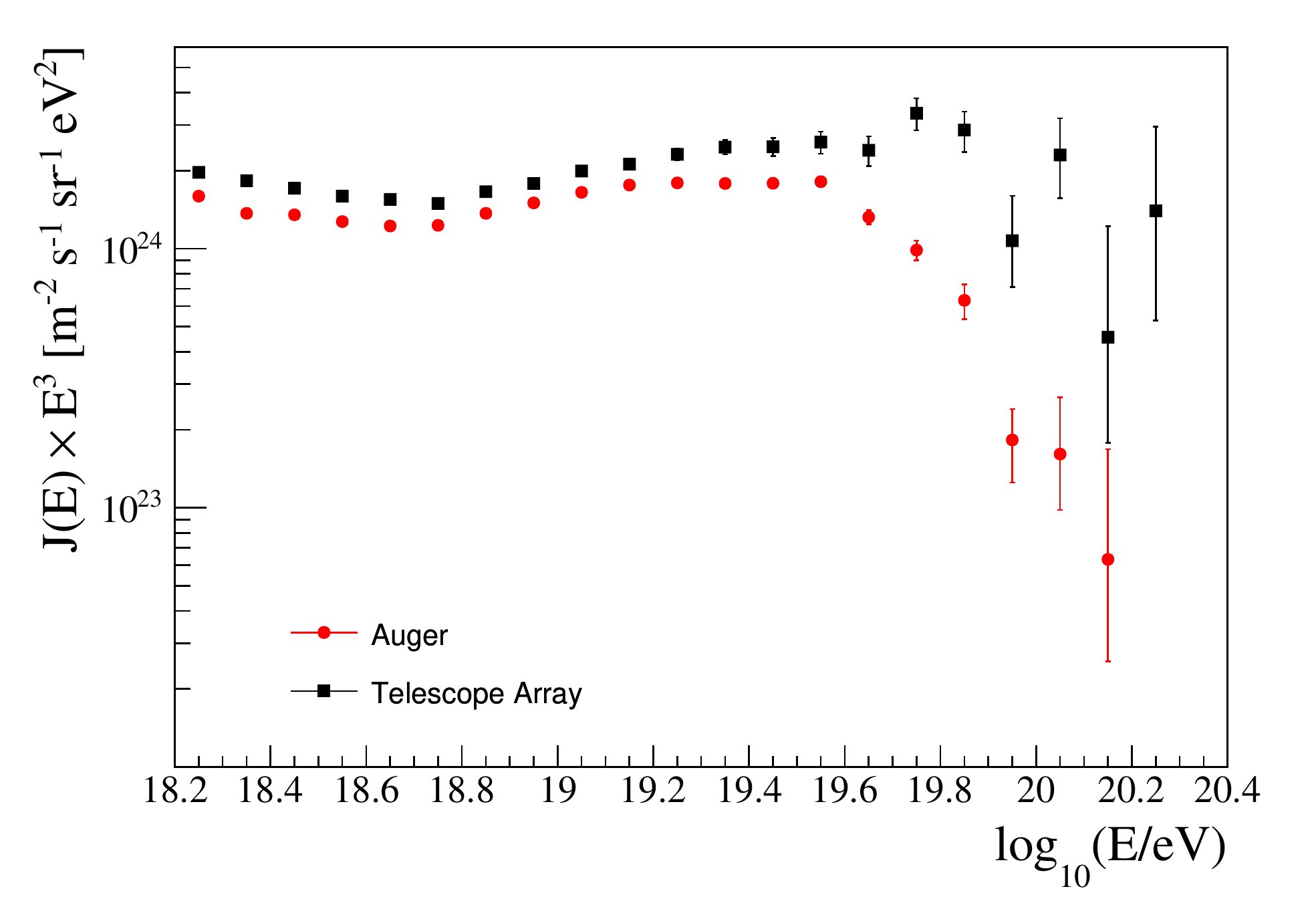}
\includegraphics[width=0.5\textwidth]{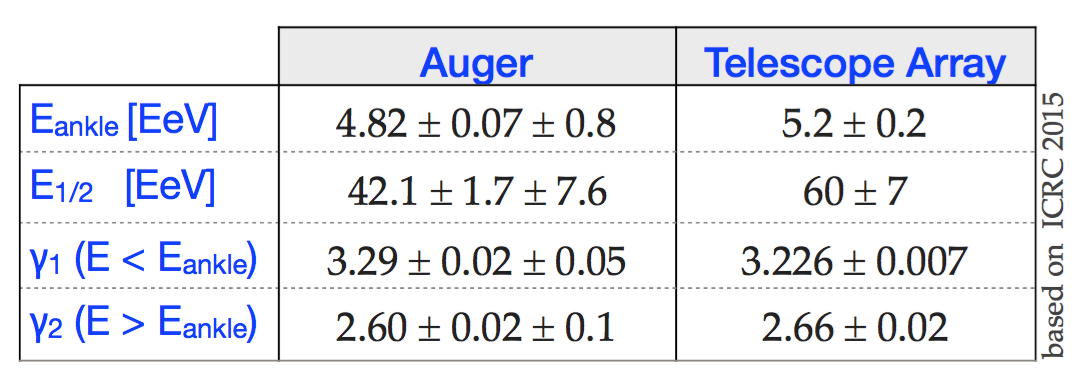}
\caption{Energy spectrum measured by the Telescope Array and Auger observatories above 10$^{18}$~eV. The measured spectral features are reported in the table~\cite{AugerSpectrum2015,TASpectrum2015}. \label{fig:spectra}}
\end{figure}

\begin{figure}[!t]
\centering
\includegraphics[width=0.48\textwidth]{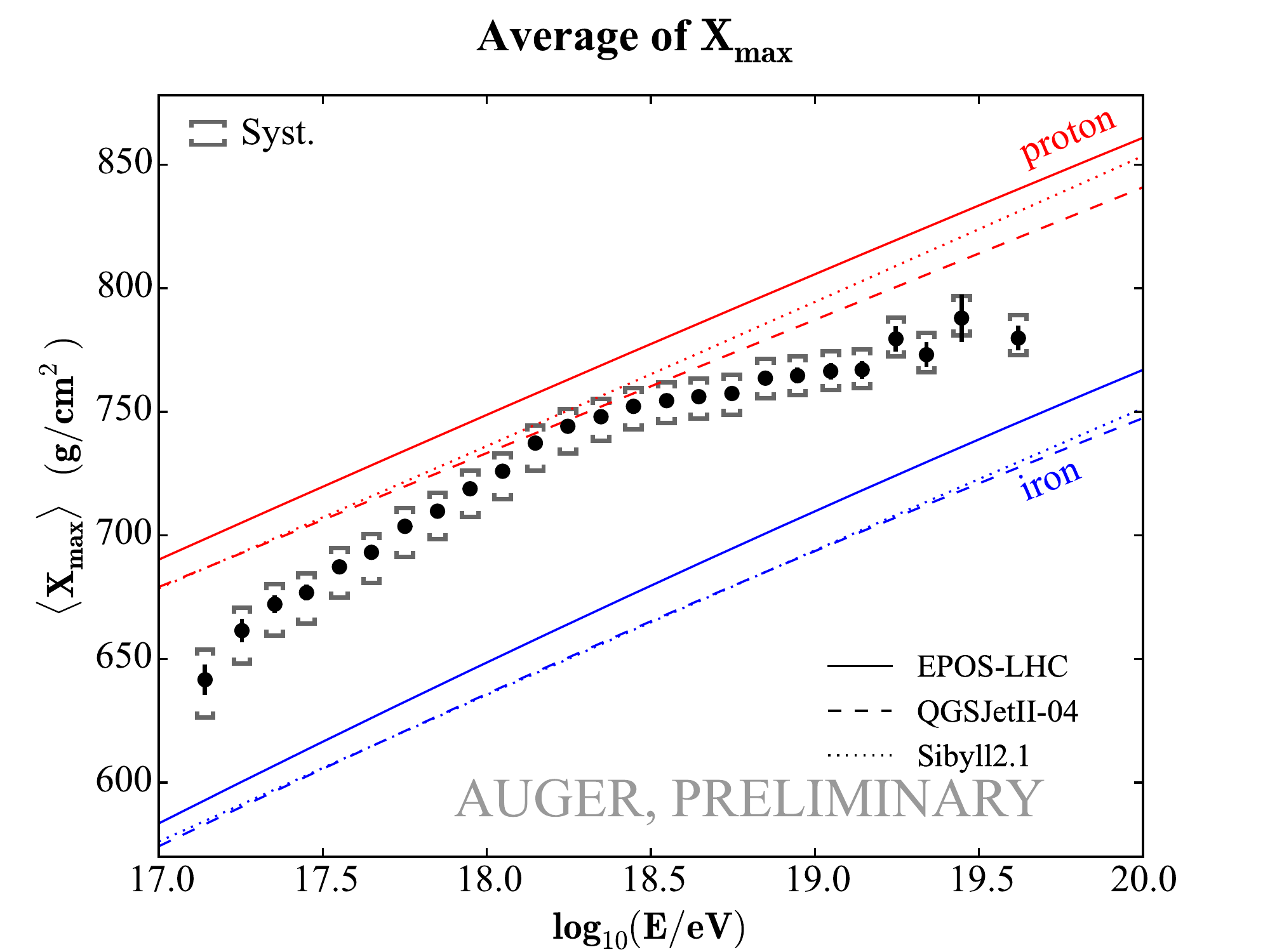}
\includegraphics[width=0.48\textwidth]{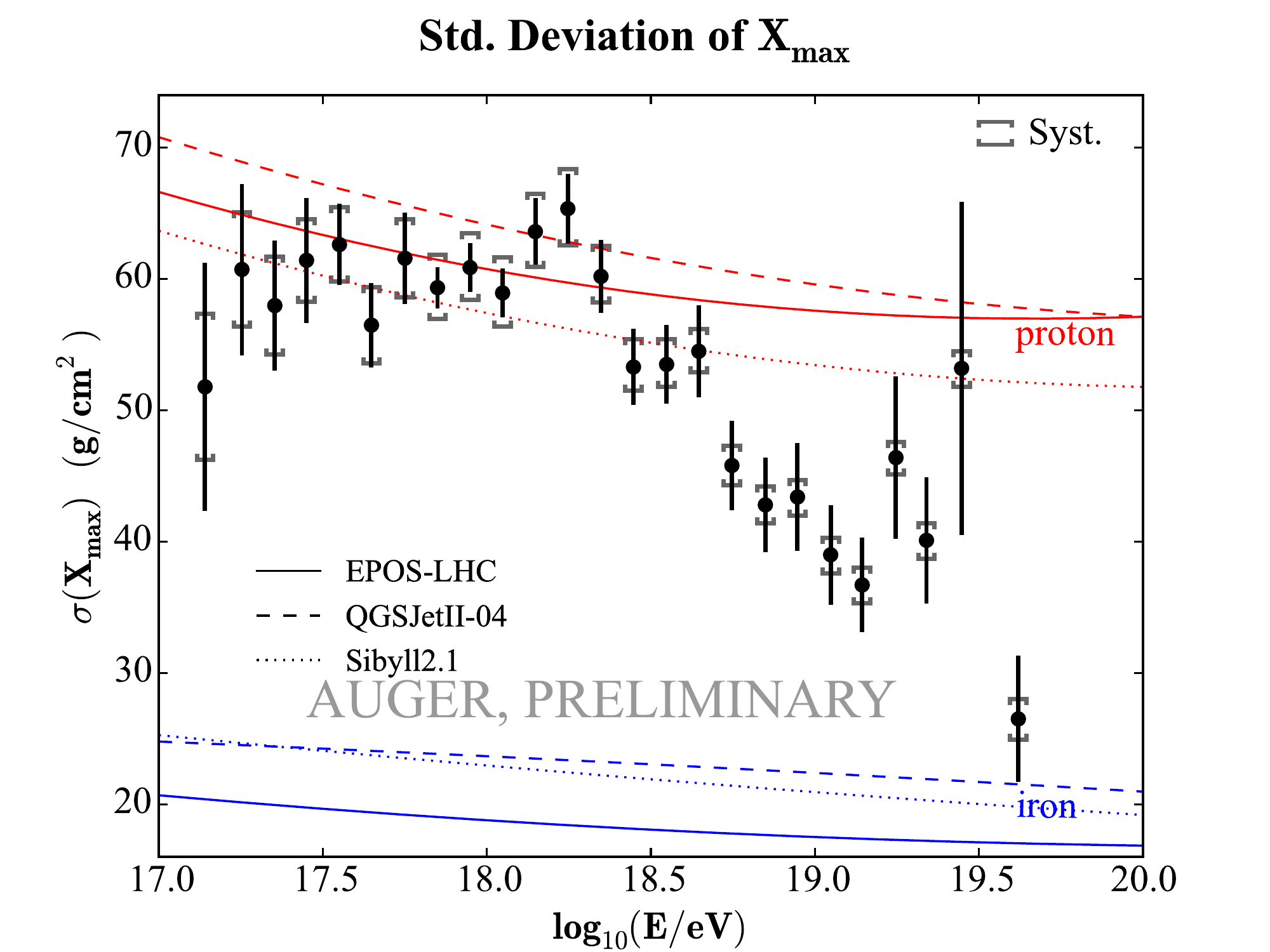}
\caption{\label{fig:Xmax}Evolution of the first two moments of the $X_{\rm max}$ distribution with energy for the Pierre Auger Observatory. Expectation for proton and iron primaries are shown as lines for different hadronic interaction models. Systematic uncertainties are shown in brackets~\cite{XmaxAugerICRC2015}. }
\end{figure}

\begin{figure}
\centering
\includegraphics[width=0.5\textwidth]{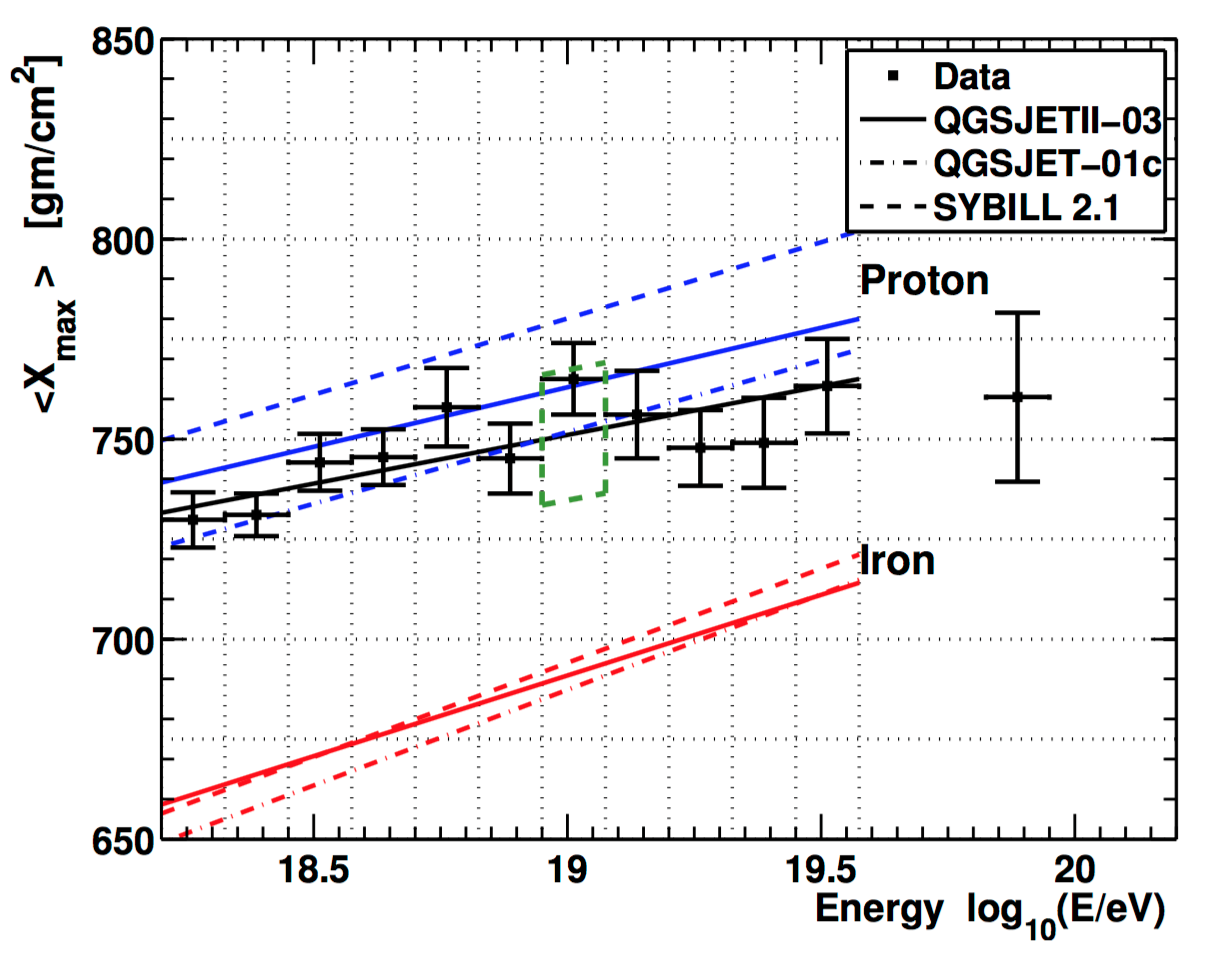}
\caption{\label{fig:XmaxTA} Evolution of $\langle\Xm\rangle$ with energy for the Telescope Array. The expectations for proton and iron primaries, shown as lines for different hadronic interaction models, are folded with the detector resolution and efficiency. Systematic uncertainties are indicated by the green box~\cite{XmaxTA,XmaxTAICRC2015}}
\end{figure}

\begin{figure}
\centering
\includegraphics[width=0.51\textwidth]{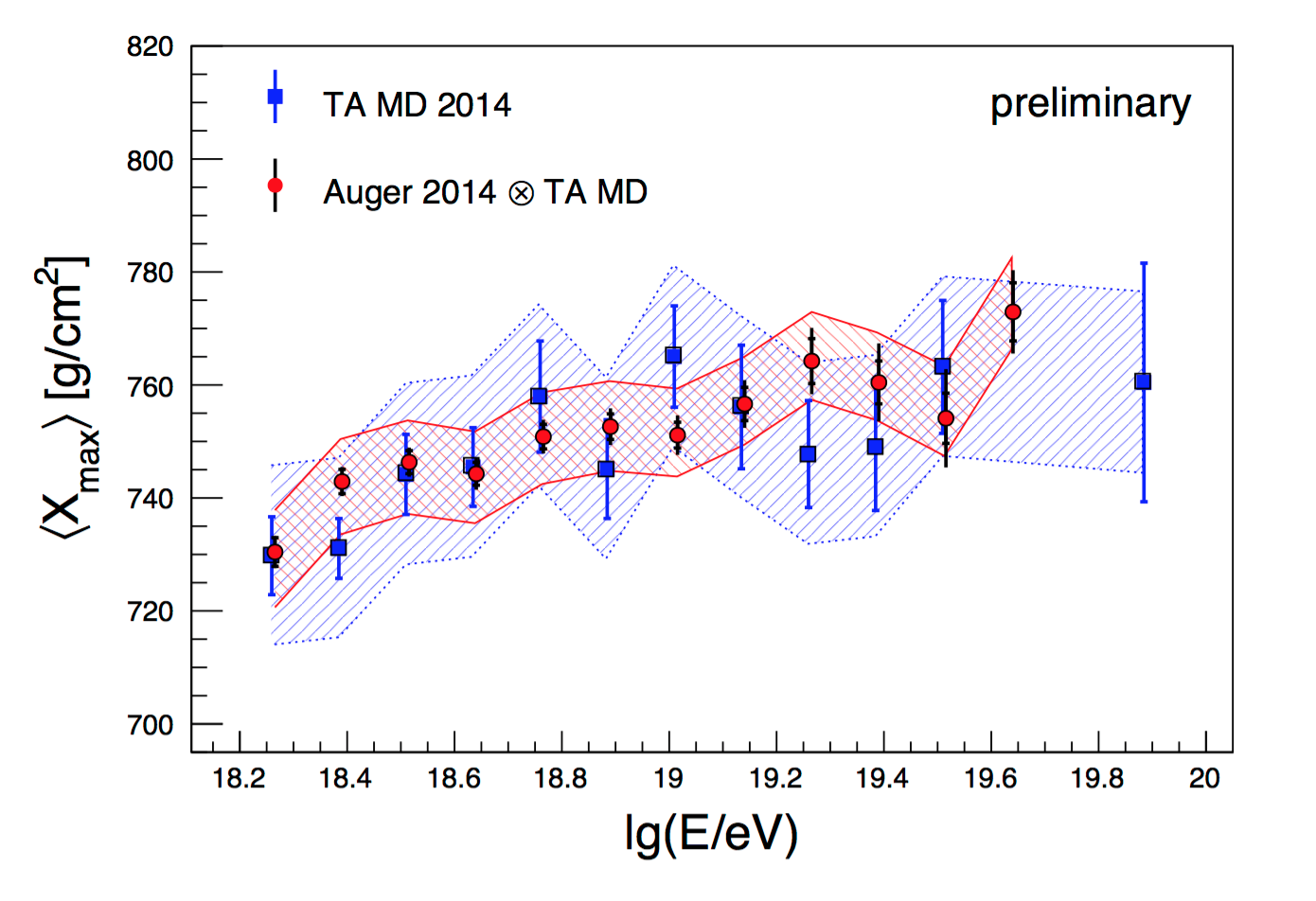}
\caption{\label{fig:XmaxAugerTA} Comparison of $\langle\Xm\rangle$ measured by Telescope Array (blue square) compared to the reconstructed $\langle\Xm\rangle$ when injecting the Auger composition mix (red circles)~\cite{UngerICRC2015}. }
\end{figure}

\section{Mass composition}

The chemical composition is inferred from the measurement of the first two moments of the $X_{\rm max}$ distribution ($\langle\Xm\rangle$ and $\sigma$($X_{\rm max}$)).  The average depth of the shower maximum, $\langle\Xm\rangle$ is proportional to the logarithm of energy and to $\langle\ln A\rangle$, with $A$ the atomic mass of the primary cosmic ray. The evolution of $\langle\Xm\rangle$ with energy, named ``elongation rate", is expected to be mostly independent of the primary particle and to be constant against energy. In Fig.~\ref{fig:Xmax} the predictions from Monte Carlo simulations are shown for protons and irons and for post-LHC hadronic models. 
In data, the \Xm is measured using high-quality-selected hybrid events. This implies that, because of the limited FD duty cycle, the mass composition measured with \Xm only extends up to 10$^{19.5}$~eV. 

The results of the Pierre Auger Observatory are shown in Fig.~\ref{fig:Xmax} for the $\langle\Xm\rangle$ (top) and $\sigma(\Xm)$ (bottom), compared to the Monte Carlo predictions.  Using the superposition model the mean logarithmic mass can be found from the measured $\langle\Xm\rangle$. The two observables suggest similar conclusions:  a composition evolving from mixed to light primaries at low energies, a break at energies of about 10$^{18.3}$~eV - interestingly close to the ankle region - and then getting heavier with increasing energy.  
The measured $\sigma$($X_{\rm max}$) indicates that composition changes from a mixture of several components to one dominated by a few element~\cite{XmaxAuger,XmaxAugerICRC2015}. 
 
An independent test of the spread of the mass composition has been performed investigating the correlation between the \Xm and the SD energy estimator $S_{\rm{ropt}}$. The correlation coefficient depends on the purity of the dataset and on the primary type (being negative in the case of a mass mixture) and is robust against assumptions on the hadronic interaction models. The analysis of events in the energy range $3-10 \times  10^{18}$~eV indicates that a pure composition or a mixture of proton and helium primaries is not compatible with data, disfavouring scenarios with a almost pure composition, as in the proton-dip model~\cite{XmaxS1000}.

Results from Telescope Array are shown in Fig.~\ref{fig:XmaxTA} for energies between 10$^{18}$~eV and 10$^{19.5}$~eV~\cite{XmaxTA,XmaxTAICRC2015}. They are interpreted by the collaboration as the indication of a pure proton composition over the full energy range. It is worthwhile to remind that the interpretation of the \Xm moments in terms of an average mass composition depends on the hadronic models which are largely unknown and are based on the extrapolation of the low energy accelerator's data. \\

The comparison of the Auger and Telescope Array results is not straightforward because of the different analysis approaches used by the two collaborations. For the Auger Observatory selection criteria are applied to the data to remove the bias due to the limited field of view of the fluorescence telescopes. Because of the strict selection, this method is not adopted by the Telescope Array Collaborations and the \Xm distribution includes detector effects (e.g., related to the acceptance and selection efficiency). To infer the mass composition, data are thus compared to simulations which have been processed through the simulation and reconstruction chains.  
With the aim of understanding the differences between Auger and Telescope Array results, a joint analysis has been performed,  reconstructing the composition mix observed by the Pierre Auger Observatory using the full analysis chain of the Telescope Array collaboration.  The reconstructed $\langle\Xm\rangle$ against energy is drawn in Fig.~\ref{fig:XmaxAugerTA} compared to the one measured by Telescope Array from data. The obtained agreement points out that the current uncertainties with the Telescope Array statistics is too large to discriminate between the Auger reconstructed mixture and a pure proton composition~\cite{UngerICRC2015}.  \\

\begin{figure}
\centering
\includegraphics[width=0.45\textwidth]{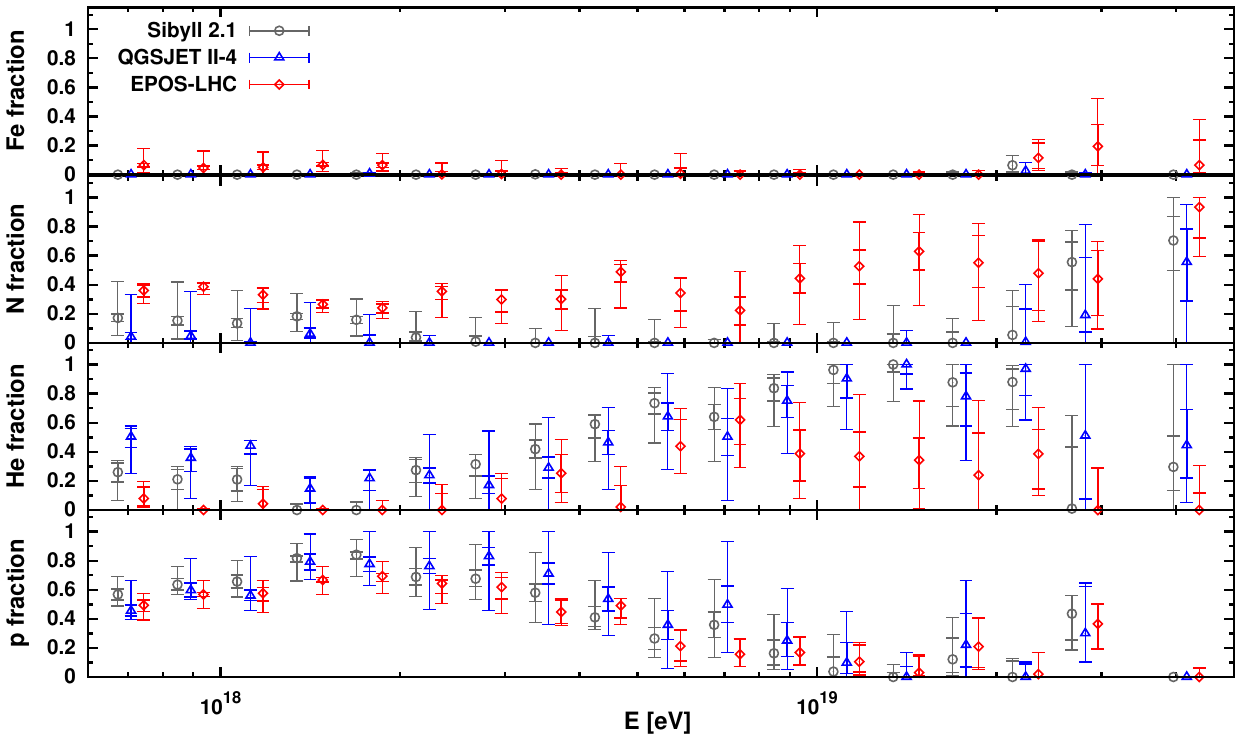}
\caption{\label{fig:Implications}Fitted fraction of a mixture of four species: proton, helium, nitrogen, iron (from bottom to top) and for different hadronic models~\cite{XmaxInterpretation}. }
\end{figure}

The Auger Collaboration has also shown~\cite{XmaxInterpretation} that the two first moments of an $X_{\rm max}$ distribution can be described with different mass composition mixtures. To avoid this degeneracy the full $X_{\rm max}$ distribution is fitted with simulation templates  assuming a mixture of $N$-components whose abundances are free parameters of the fit. The best description of the data is obtained with four components (proton, helium, nitrogen and iron nuclei).  All the models predict a similar behaviour with a large fraction of protons at energies around the ankle and a sub-dominant iron content over the full energy range (Fig~\ref{fig:Implications}). \\
 
 A combined fit of the energy spectrum and the mass composition measured by the Pierre Auger Observatory~\cite{AugerNIM2015} - under simple assumptions on the astrophysical sources and on the propagation of cosmic rays - seems to favour a scenarios with a limitation of the maximum acceleration at the source~\cite{AugerFluxInterpretation}. On the other hand, the fit of the energy spectrum measured by the Telescope Array Collaboration, assuming a pure proton composition at the source, constrains the source properties and prefers a GZK scenario with a strong source evolution~\cite{TAFluxInterpretation}. This scenario is however challenged by the recent limits on cosmogenic neutrino fluxes and by the diffuse sub-TeV $\gamma$-radiation (see e.g., ~\cite{Heinze:2015hhp, Aartsen:2016ngq, AugerNeutrino, Berezinsky2016}).

\section{Hadronic interactions and muon content}

\begin{figure}
\centering
\includegraphics[width=0.48\textwidth]{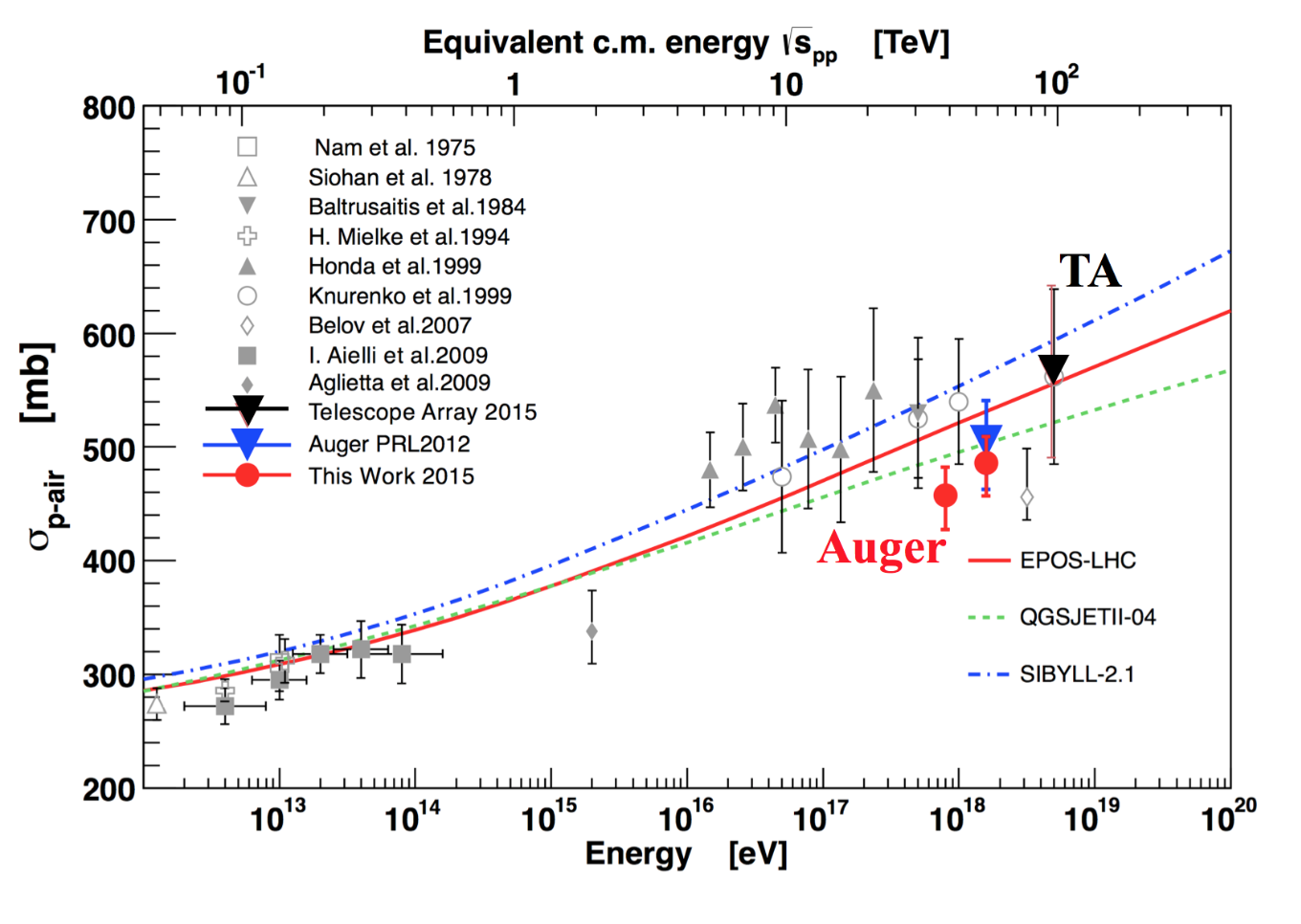}
\caption{Measurement of the proton-air collision cross section from the Telescope Array and Auger observatories~\cite{VerziICRC2015}. \label{fig:hadronic}}
\end{figure}

In connection with the mass composition interpretation, it is worthwhile to mention the possibility to test hadronic physics with air-showers at center-of-mass energies that are one or two orders of magnitude higher than the ones reached at LHC. A measurement of the proton-air cross-section has been performed by the Pierre Auger and Telescope Array collaborations from the fit of the tail of the $X_{\rm max}$ distribution for a  sample of proton-dominated events. To select a proton-dominated dataset the Auger collaboration uses an energy range around 10$^{18.5}$~eV whereas for Telescope Array the full energy range is used. A detailed description of the two analyses is given in~\cite{xsecAuger,xsecTA}. The proton-air cross-section $\sigma_{\rm{p-Air}}$ is shown in Fig.~\ref{fig:hadronic}, compared to predictions from the most up-to-date hadronic interaction models. 

 \begin{figure}
\centering
\includegraphics[width=0.43\textwidth]{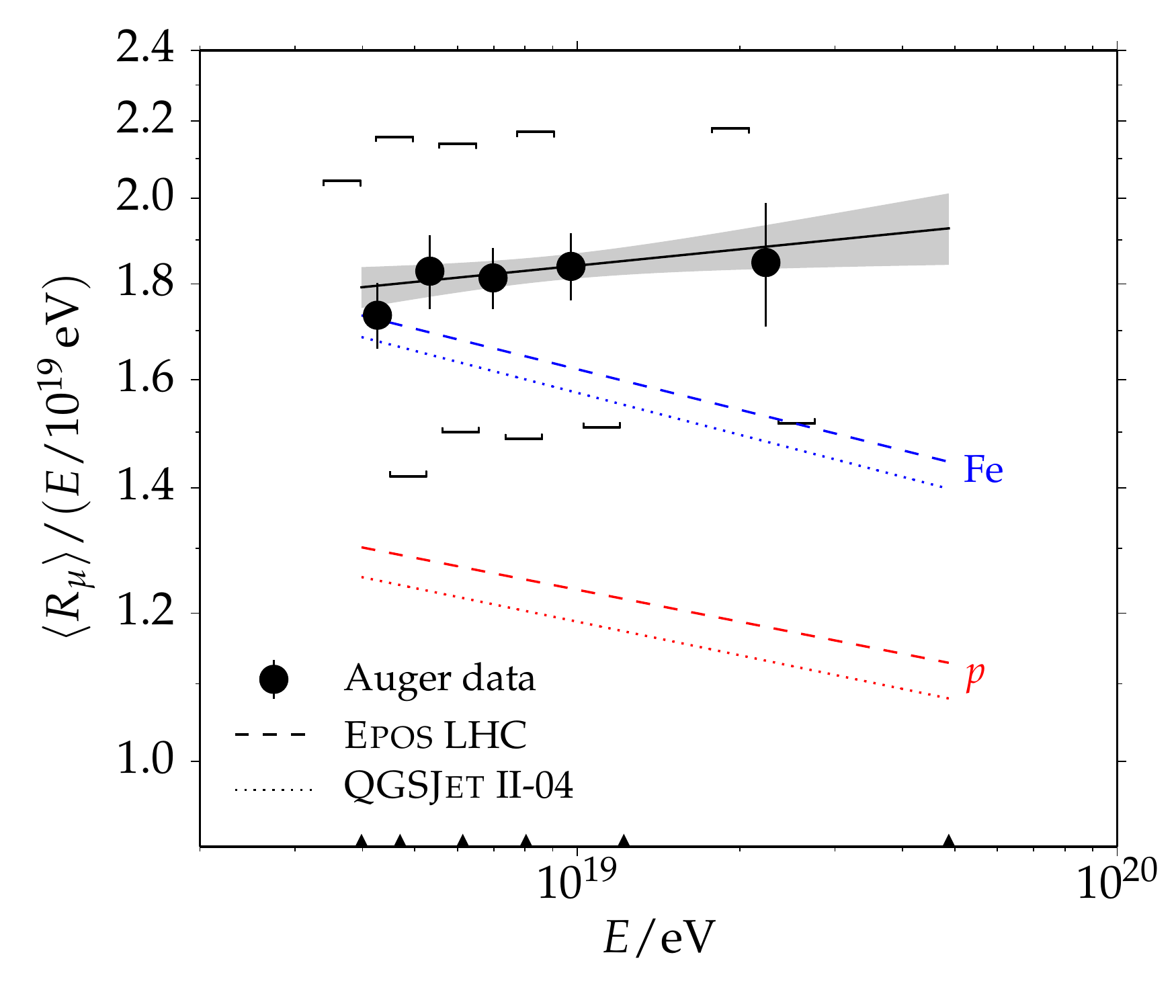}
\includegraphics[width=0.4\textwidth]{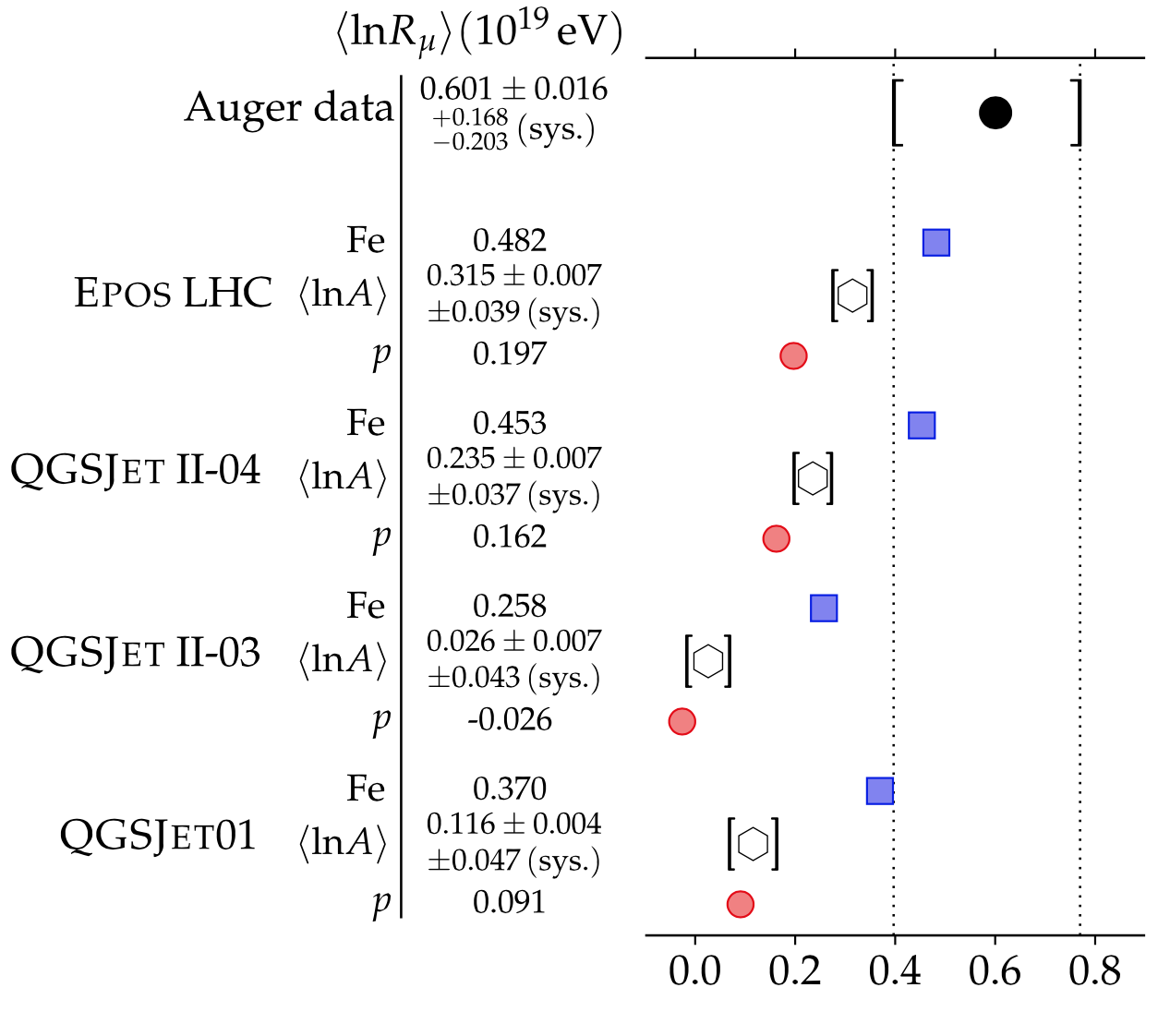}
\caption{\label{fig:Nmu} Top: Muon number estimator $\langle R_\mu \rangle$ as a function of energy from data and simulations. Systematic uncertainties indicated by brackets. Bottom: muon deficit observed comparing air-shower simulations and Auger data for different hadronic models and mass composition assumptions~\cite{MuonContent}. }
\end{figure}

The water-Cherenkov stations of the Pierre Auger Observatory are sensitive to the muonic and electromagnetic component. Even if in the current design these two components cannot be accessed separately~\cite{LSD,AugerPrime}, an estimate of the muon content in the air shower is possible by selecting events with large zenith angles, for which the atmosphere acts as a shield of the electromagnetic component. 
A parameter, $R_{\mu}$, is introduced as the ratio of the measured number of muons and the expected value for a reference model. The separation between the expectations for proton and iron induced showers proves the power of $R_{\mu}$ as a composition estimator. As shown in Fig.~\ref{fig:Nmu}, the measured muon number, higher than for the pure iron case, is not compatible with simulations and suggests a muon deficit in simulations, varying between 30 and 80\% depending on the hadronic model~\cite{MuonContent}. 
This result is also confirmed by an independent analysis using hybrid events with energy between 6 and 16$\times10^{18}$~eV and zenith angle smaller than 60$^\circ$. 
A set of proton and iron simulated air-showers matching the longitudinal profile in data is produced for different hadronic interaction models. The signal recorded in data is larger than simulated one. A model is fit to data with a rescaling factor for the electromagnetic and the hadronic components. Whereas the rescaling factor for the electromagnetic component is close to unity within the uncertainty, an excess is found for the hadronic component ranging between 30 and 60\% for EPOS-LHC and QGSJET-II.04 models, respectively~\cite{Farrar}. \\

\begin{figure}
\centering
\includegraphics[width=0.5\textwidth]{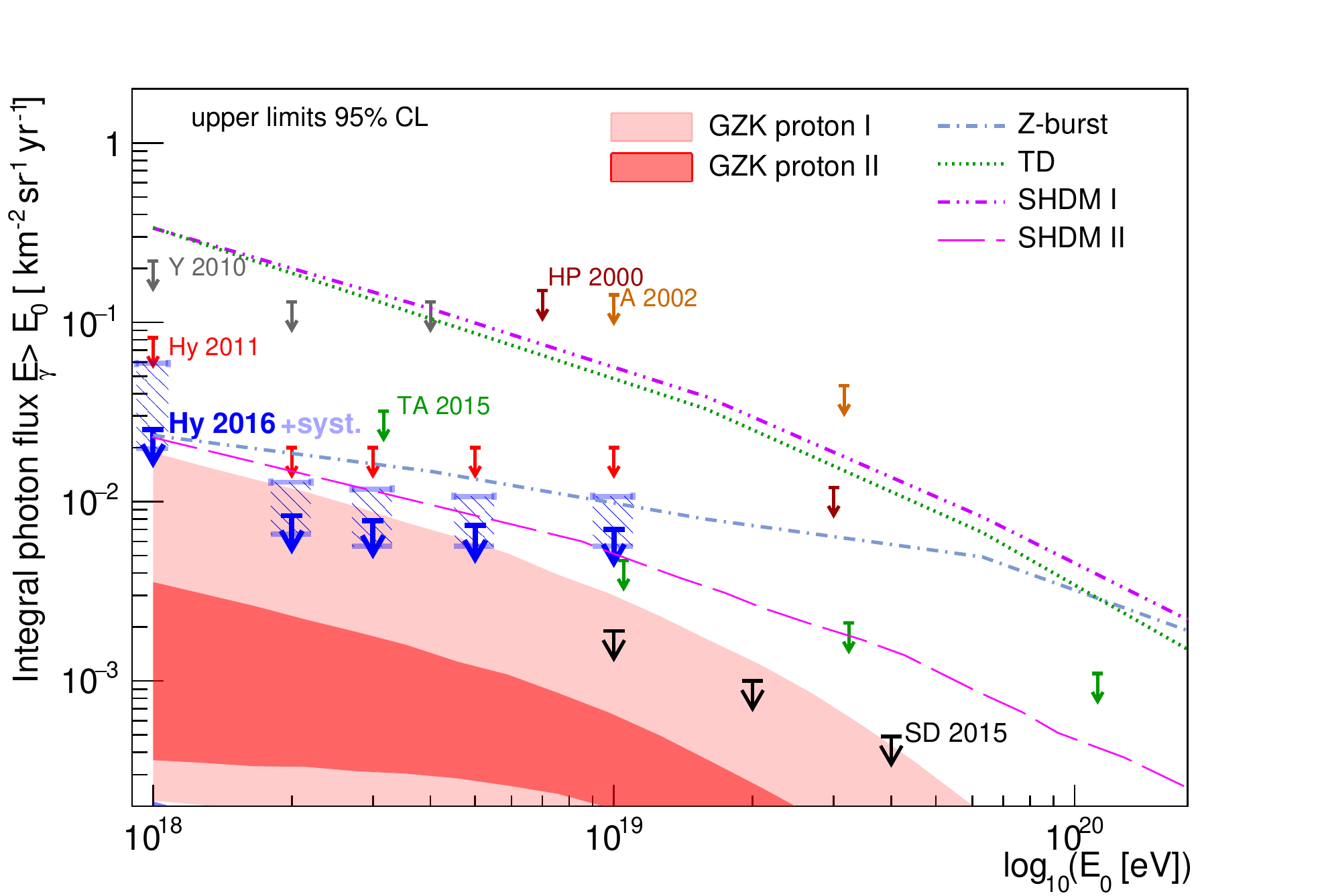}
\caption{Upper limits on the diffuse integral photon flux for several experiments. The expectation for exotic models and in the propagation scenario (GZK) are shown for comparison~\cite{AugerPhotons2016}. \label{fig:photons}}
\end{figure}

\section{Cosmogenic photons and neutrinos}

The UHE photons and neutrinos are specific signature of the GZK process.  A flux of photons is also expected in several top-down models in which UHECRs are the secondary products of the decay of super-massive particles. Neutrinos have the advantage of travelling cosmological distances without interacting and they can thus trace back their production sites. On the contrary, photons undergo interactions with the extragalactic background light inducing electromagnetic cascades~\cite{EleCa}. The expected flux of cosmogenic photons and neutrinos are thus sensitive to several astrophysical parameters, as the source properties (spectral shape, maximum acceleration energy, distribution and cosmological evolution, chemical composition) and the extragalactic ambient (e.g., background light and magnetic fields).  

The search for photons and neutrinos above 10$^{18}$~eV with ground-based experiments for UHECRs profit from the different shower development with respect to hadron-induced air-showers.  
Given their mostly electromagnetic nature, on average, photon-induced air showers develop deeper in the atmosphere, compared to hadronic ones of similar energies. In addition, a significantly smaller muon content, compared to hadron induced showers, is expected. The search for UHE photons is conducted by Telescope Array and the Auger Observatory by means of the lower signal strength and steeper LDF shape at the ground~\cite{TAPhotonsICRC2015,AugerPhotonsICRC2015}. No photon-like events have been identified and upper limits on the integral photon flux have been set. They are shown in Fig.~\ref{fig:photons} compared to some model predictions. Several top-down models are ruled-out or constrained and the current limits obtained by the Auger Collaboration are below the most optimistic ``propagation scenarios" for proton sources. The expected flux for iron or a mixed composition is up to a factor hundred lower than protons and, incoming years, it can be reached in some configurations, by increasing the exposure and improving the background discrimination capabilities of the current detectors~\cite{AugerPhotons2016}. \\

\begin{figure}
\centering
\includegraphics[width=0.5\textwidth]{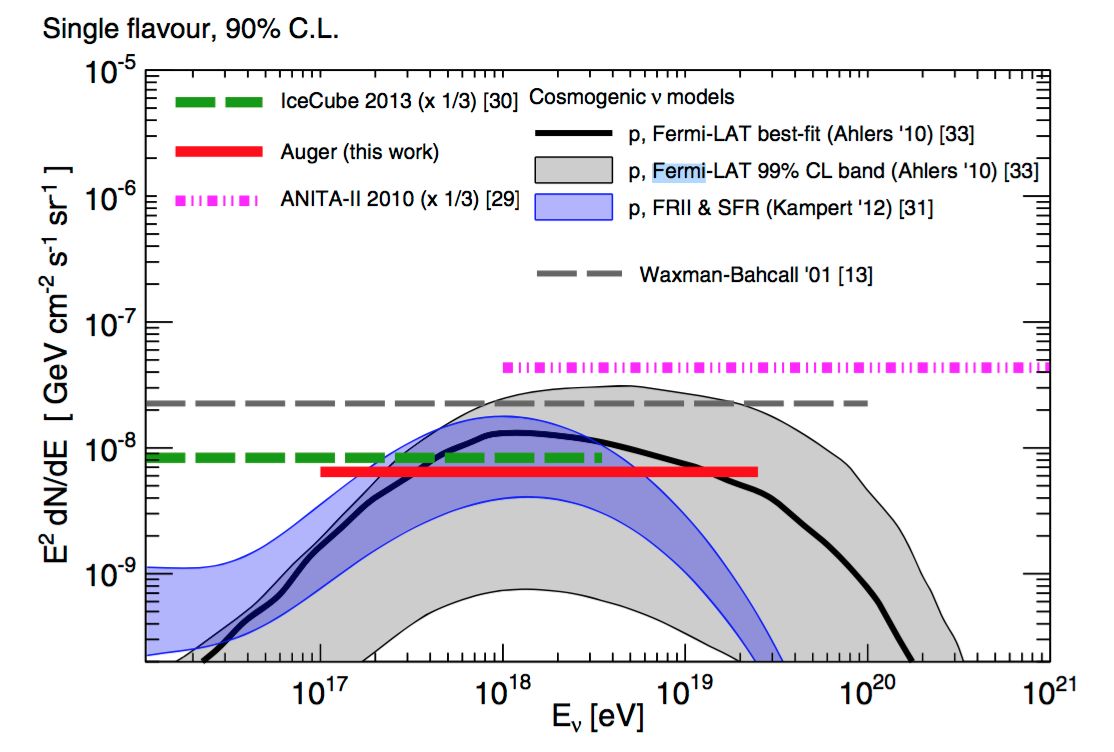} 
\caption{\label{fig:neutrinos}Upper limits (at 90\% C.L.) to the normalization of the diffuse flux of UHE neutrinos, compared to expectation from cosmogenic neutrinos in a pure proton scenario~\cite{AugerNeutrino}. }
\end{figure}

In contrast to nuclei and photons, neutrinos initiate showers close to the ground level. The electromagnetic component of hadron-induced air-showers with large zenith angles, is mostly absorbed in atmosphere and the shower front at ground level is thus dominated by muons. On the contrary, cascades induced by neutrinos develop deep in the atmosphere and have a substantial electromagnetic component at the ground.  
Given its design, the water-Cherenkov detectors offer a significant exposure to almost-horizontal events allowing to search for  neutrinos by selecting inclined (or Earth-skimming) events with a large electromagnetic component~\cite{AugerNeutrino}. The selection criteria are tuned to ensure a very low background contamination,  less than 1 expected event per 50 yr on the full SD array. 
Upper limits on the normalization factor for a differential neutrino flux proportional to E$^{-2}$ are derived at 90\% confidence level (Fig.~\ref{fig:neutrinos}). These limits disfavour cosmogenic models with a pure proton composition at the sources and a strong evolution of the sources as well as models of proton sources with a GeV $\gamma$-ray flux constrained by Fermi-LAT data. 

\section{Summary and future perspectives}

The extremely low fluxes of extragalactic cosmic-rays makes challenging the study of their origin and their nature. In the past decade, the improvements of the FD and SD detection techniques, the conception of the hybrid operation mode to remove the dependence on the air-showers models and the construction of observatories with huge exposures has allowed to obtain fundamental results, as the precise determination of the energy spectrum, the mass composition measurement and the possibility to test hadronic interaction models in an energy regime not-accessible with accelerators. Nevertheless some results are unexpected and still controversial, in particular the weak anisotropy signal at ultra-high energies, the interpretation of the flux cut-off and of the \Xm data, the muon deficit observed in simulations. The mass composition in the flux suppression region, currently unexplored because of the FD limited duty cycle, is one of the observables that can play a crucial role to discriminate between different scenarios and to provide hints on particle physics.  

With the necessity of increasing the statistics and the available information on extensive air-showers, the Telescope Array and the Auger Observatory foresee to upgrade of their detectors. The Telescope Array plans to extend its surface by a factor four (``TAx4"), with the aim of a significant increase in the event statistics at the energies~\cite{TAx4}. This extension will allow a better determination of the flux suppression and to test with high significance the observed hot-spot (currently at 3.4 $\sigma$, on an angular scale of 20$^\circ$)~\cite{TAHotSpot}. 
The AugerPrime upgrade mostly consists in the installation of a 4~m$^2$ scintillator on top of each surface station, in the upgrade of the electronics and in the completion of the underground muon detector~\cite{AugerPrime}. The main goal is to have a separate measurement of the electromagnetic and muonic component at the ground in order to perform a mass composition study using the surface detector (duty cycle about 10 times larger than the FD one). The anisotropy can then be tested by selecting the lightest observed component and doing astronomy with UHECR may be explored if at least 10\% of protons are found in data. Finally, the measured muon content will also add constrains to the hadronic interaction models.
Both these upgrades are expected to be operational in a few years. 

\bigskip % extra skip inserted
\begin{acknowledgments}
The work of M.S. made in the ILP LABEX (ANR-10-LABX-63), is supported by French state funds managed by the ANR within the Investissements d'Avenir programme (ANR-11-IDEX-0004-02). 
\end{acknowledgments}

\bigskip % extra skip inserted

\end{document}